\documentclass[aps, prb, twocolumn, superscriptaddress, showpacs]{revtex4}
\usepackage{mathrsfs, amsfonts, amsmath, amssymb, graphicx, bm, epsfig}
\usepackage[colorlinks, linkcolor=blue, anchorcolor=blue, citecolor=red]{hyperref}

\begin{document}
\title{Microscopic derivation of spin-transfer torque in ferromagnets}
\author{Ran Cheng}
\email{rancheng@utexas.edu}
\affiliation{Department of Physics, University of Texas at Austin, Austin, Texas 78712, USA}
\author{Qian Niu}
\affiliation{Department of Physics, University of Texas at Austin, Austin, Texas 78712, USA}
\affiliation{International Center for Quantum Materials, Peking University, Beijing 100871, China}

\begin{abstract}
 Spin-transfer torque (STT) provides key mechanisms for current-induced phenomena in ferromagnets. While it is widely accepted that STT involves both adiabatic and non-adiabatic contributions, their underlying physics and range of validity are quite controversial. By computing microscopically the response of conduction electron spins to a time varying and spatially inhomogeneous magnetic background, we derive the adiabatic and non-adiabatic STT in a unified fashion. Our result confirms the macroscopic theory [Phys.~Rev.~Lett.~\textbf{93},~127204~(2004)] with all coefficients matched exactly. Our derivation also reveals a benchmark on the validity of the result, which is used to explain three recent measurements of the non-adiabatic STT in quite different settings.
\end{abstract}

\maketitle

The interplay between current and magnetization is currently the central topic of spintronics~\cite{ref:Spintronics}. When a current flows through a ferromagnetic metal, it becomes spin-polarized due to local exchange coupling between conduction electron spins and local magnetic moments. In turn, spin angular momentum is transferred to magnetization through the mechanism known as spin-transfer torque (STT)~\cite{ref:BegerSlonczewski,ref:STT}, which is a consequence of spin conservation. STT provides key mechanisms for numerous intriguing phenomena in ferromagnets, such as current-driven domain wall motion~\cite{ref:DWDynamics1,ref:DWDynamics2}, spin wave excitations~\cite{ref:SW1,ref:SW2}, \emph{etc}. In both fundamental studies and device designs, STT-driven magnetization dynamics has aroused enormous attention in the past two decades~\cite{ref:STTReview1,ref:STTReview2}, and it is becoming the core issue of spintronics. However, the fundamental physics underlying STT is far from clear.

At present, STT is believed to be divided into adiabatic (reactive) and non-adiabatic (dissipative) contributions. While the former has been derived microscopically via different approaches~\cite{ref:BegerSlonczewski,ref:AdiabaticSTT1}, the latter has only been justified macroscopically through spin conservation~\cite{ref:STT,ref:Fitting,ref:SpinMHD} and Galilean invariance~\cite{ref:GalileonInv}, whose microscopic origin is under intense debates. In many recent efforts, microscopic theories have been developed in generic ways~\cite{ref:NonAdiabaticSTT1,ref:NonAdiabaticSTT2,ref:NonAdiabaticSTT3,ref:NonAdiabaticSTT4,ref:NonAdiabaticSTT5} and in specific contexts~\cite{ref:MomentumTransfer,ref:EuropModel,ref:ClassicalModel,ref:DWSTTModel}, but their coefficients do not lead to a consensus. Meanwhile, some others even cast doubt on the existence of the non-adiabatic STT~\cite{ref:CastDoubt}. From an experimental point of view, measurements of this torque are not in agreement~\cite{ref:Exp1,ref:Exp2,ref:NarrowDW}, and the magnitude is sensitive to spin-orbit interaction~\cite{ref:Miron} and impurity doping~\cite{ref:Lepadatu}.

In this Letter, a microscopic derivation of the magnetization dynamics induced by STT is provided. Based on the \emph{s-d} model~\cite{ref:STT}, we first calculate the response of a conduction electron spin to a time varying and spatially inhomogeneous magnetic background $\bm{M}(\bm{r},t)$, and we obtain the non-equilibrium local spin accumulation $\delta\bm{m}$ (perpendicular to $\bm{M}(\bm{r},t)$) by integration over the conduction band. Due to the exchange coupling between \emph{s}-band electrons and \emph{d}-band magnetic moments, the back-action exerted on $\bm{M}(\bm{r},t)$ by the current is proportional to $\delta\bm{m}\times\bm{M}$, where the adiabatic and non-adiabatic STTs naturally appear on an equal footing. Our result (Equation~\eqref{eq:finalresult}) justifies the macroscopic model~\cite{ref:STT} with all coefficients matched exactly. Our derivation also provides a benchmark on the validity of the result, which is used to explain three experimental results: why the non-adiabatic STT on narrow domain walls~\cite{ref:NarrowDW} shows deviations from Eq.~\eqref{eq:finalresult}, why Eq.~\eqref{eq:finalresult} is still valid even when an extraordinarily large non-adiabatic STT is achieved~\cite{ref:Miron}, and why the non-adiabatic STT is enhanced by impurity doping while the damping is not affected~\cite{ref:Lepadatu}.

We adopt the \emph{s-d} model where electron transport is due to the itinerant \emph{s}-band. It will be treated separately from the magnetization, which mostly originates from the localized \emph{d}-band. The conduction electrons interact with the magnetization through the exchange coupling described by the following Hamiltonian
\begin{align}
 H_{ex}=\frac{SJ_{ex}}{M_s}\bm{s}\cdot\bm{M}(\bm{r},t) \label{eq:Hamiltonian}:
\end{align}
where $\bm{s}$ is the (dimensionless) spin of a conduction electron, $|\bm{M}(\bm{r},t)|=M_s$ is the saturation magnetization, and $S$ denotes the magnitude of background spins. The coupling strength $J_{ex}$ can be as large as an $\mathrm{eV}$ in transition metals and their alloys, so that if $\bm{M}(\bm{r},t)$ varies slowly in space and time, conduction electron spins will follow the background profile when the system is in thermal equilibrium, which is known as the adiabatic limit. However, when an external current is applied to the system, a small non-equilibrium spin accumulation $\delta\bm{m}$ transverse to local $\bm{M}(\bm{r},t)$ is induced. It is this $\delta\bm{m}$ that exerts STT on the background magnetization.

\begin{figure}[t]
   \centering
   \includegraphics[width=0.95\linewidth]{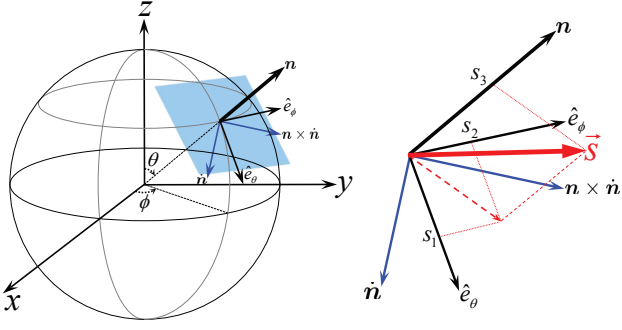}
   \caption{(Color online) Eigenstates of Eq.~\eqref{eq:Hamiltonian} form a set of local spin bases, and define a local frame that moves with $\bm{n}=\bm{M}/M_s$. Components of the conduction electron spin $\bm{s}$ (red) in the local frame are denoted by $s_1$, $s_2$, and $s_3$. In the tangential plane with normal $\bm{n}$, we make a coordinate transformation from $\hat{e}_\theta$ and $\hat{e}_\phi$ to $\dot{\bm{n}}$ and $\bm{n}\times\dot{\bm{n}}$ so that everything expressed in the new basis is physical. \label{Fig:Spin}}
\end{figure}

To compute $\delta\bm{m}$, we first study the spin response of an individual conduction electron to the background $\bm{M}(\bm{r},t)$ when current is applied. From Eq.~\eqref{eq:Hamiltonian}, we know that \textit{local} spin-up (majority) and spin-down (minority) bands are separated by a large gap $\Delta\equiv SJ_{ex} =\mathcal{E}_{\downarrow}-\mathcal{E}_{\uparrow}$, and the associated spin wave functions are denoted by $|\!\uparrow\!(\bm{r},t)\rangle$ and $|\!\downarrow\!(\bm{r},t)\rangle$. The electron is described by a coherent wave packet centered at $(\bm{r}_c,\bm{k}_c)$~\cite{ref:Niu,ref:Ran}
\begin{align}
 |W\rangle=\!\int\mathrm{d}^3\bm{k}\ w(\bm{k})e^{i\bm{k}\cdot\bm{r}}|\bm{k}\rangle[c_a|\!\uparrow\!(\bm{r}_c,t)\rangle+c_b|\!\downarrow\!(\bm{r}_c,t)\rangle],
\end{align}
where $w(\bm{k})$ is a profile function that satisfies $\int\mathrm{d}\bm{k}\bm{k}|w(\bm{k})|^2=\bm{k}_c$; $|\bm{k}\rangle$ is the periodic part of the \textit{local} Bloch function; and $c_a$, $c_b$ are superposition coefficients. Since $|\!\uparrow\!(\bm{r}_c,t)\rangle$ and $|\!\downarrow\!(\bm{r}_c,t)\rangle$ form a set of local spin bases with the quantization axis being $\bm{n}(\bm{r}_c,t)=\bm{M}(\bm{r}_c,t)/M_s$,  we can construct a local frame moving with $\bm{M}(\bm{r}_c,t)$, where the coordinates are labeled by $\bm{n}$, $\hat{e}_{\theta}$, and $\hat{e}_{\phi}$ in Fig.~\ref{Fig:Spin}. The electron spin expressed in this local frame reads
\begin{align}
 \bm{s}&=\{s_1, s_2, s_3\}=\eta^{\dagger}\bm{\sigma}\eta \notag\\
 &=\{2\mathrm{Re}(c_a c_b^*),\ -2\mathrm{Im}(c_a c_b^*),\ |c_a|^2-|c_b|^2\}, \label{eq:localspin}
\end{align}
where $\bm{\sigma}$ is a vector of Pauli matrices, and $\eta=[c_a,c_b]^{\mathrm{T}}$ is regarded as the spin wave function in the local basis.

The equations of motion are obtained from the universal Lagrangian $L=\langle W|i\hbar\partial_t-H|W\rangle$ through the variational principle~\cite{ref:Niu}, which involves not only the dynamics of $\bm{r}_c$ and $\bm{k}_c$, but also the dynamics between the two (well separated) spin bands. The latter represents spin evolution with respect to the local magnetization $\bm{M}(\bm{r},t)$ and exhibits fast rotating character due to the large gap $\Delta$. It should be distinguished with adiabatic dynamics between degenerate bands~\cite{ref:Ran}. Due to the space-time dependence of the local spin basis, Berry gauge connections are induced in the effective Lagrangian (Appendix A)
\begin{align}
 L=i\hbar\eta^{\dagger}\dot{\eta}+\eta^\dagger[\dot{\bm{r}}_c\!\cdot\!\bm{A}+\Phi]\eta+\hbar\bm{k}_c\!\cdot\!\dot{\bm{r}}_c-\frac12s_3\Delta-\mathcal{E}_0, \label{eq:Lagrangian}
\end{align}
where $\mathcal{E}_0=\frac12(\mathcal{E}_{\uparrow}(\bm{k}_c)+\mathcal{E}_{\downarrow}(\bm{k}_c))$ denotes the average value of local band energy; the gap $\Delta$ couples only with $s_3$, which resembles a \textit{local} Zeeman energy. The Berry connections have both a spatial component,
\begin{align}
 \bm{A}=i\hbar
 \begin{bmatrix}
 \langle\ \uparrow|\nabla|\uparrow\ \rangle && \langle\ \uparrow|\nabla|\downarrow\ \rangle \\
 \langle\ \downarrow|\nabla|\uparrow\ \rangle && \langle\ \downarrow|\nabla|\downarrow\ \rangle
 \end{bmatrix}
\end{align}
as a vector potential (note that $\nabla=\frac{\partial}{\partial\bm{r}_c}$), and temporal component,
\begin{align}
 \Phi=i\hbar
 \begin{bmatrix}
 \langle\ \uparrow|\partial_t|\uparrow\ \rangle && \langle\ \uparrow|\partial_t|\downarrow\ \rangle \\
 \langle\ \downarrow|\partial_t|\uparrow\ \rangle && \langle\ \downarrow|\partial_t|\downarrow\ \rangle
 \end{bmatrix}
\end{align}
as a scalar potential. The local spin wave functions are taken to be $|\!\uparrow\rangle=[e^{-i\frac{\phi}2}\cos\frac{\theta}2, \ e^{i\frac{\phi}2}\sin\frac{\theta}2]^\mathrm{T}$ and $|\!\downarrow\rangle=[-e^{-i\frac{\phi}2}\sin\frac{\theta}2, \ e^{i\frac{\phi}2}\cos\frac{\theta}2]^\mathrm{T}$, where $\theta=\theta(\bm{r}_c,t)$ and $\phi=\phi(\bm{r}_c,t)$ are spherical angles specifying the direction of $\bm{M}(\bm{r}_c,t)$, whose total time derivatives are $\dot{\theta}=\dot{\bm{r}}_c\cdot\nabla\theta+\partial_t\theta$ and $\dot{\phi}=\dot{\bm{r}}_c\cdot\nabla\phi+\partial_t\phi$. Then the Berry connection terms can be unified into a $2\times2$ matrix
\begin{align}
 \dot{\bm{r}}_c\!\cdot\!\bm{A}+\Phi=\frac{\hbar}2
 \begin{bmatrix}
 \cos\theta\dot{\phi} && -\sin\theta\dot{\phi}-i\dot{\theta}\ \\
 -\sin\theta\dot{\phi}+i\dot{\theta} && -\cos\theta\dot{\phi}
 \end{bmatrix}.
\end{align}
It is worth mentioning that freedom exists in the choice of local spin wave functions, which leads to the gauge freedom of the Berry gauge connection. More graphically, a specified set of spin wave functions corresponds to a particular choice of local frame in Fig.~\ref{Fig:Spin}, and the relative orientation of the local frame can be rotated about $\bm{n}$ by gauge transformations, thus is not physical. But everything will be expressed in terms of gauge invariant quantities in the end.

Regarding Eq.~\eqref{eq:localspin}, the spin dynamics is obtained through the variational principle $\delta L/\delta\eta=0$. After some manipulations (Appendix B) we obtain
\begin{align}
 \begin{bmatrix}
  \dot{s}_1\\ \dot{s}_2\\ \dot{s}_3
 \end{bmatrix}
\!=\!\begin{bmatrix}
 0 & \cos\theta\dot{\phi}-\dfrac1{\tau_{_{ex}}} & -\dot{\theta} \\
 -\cos\theta\dot{\phi}+\dfrac1{\tau_{_{ex}}} & 0 & -\sin\theta\dot{\phi} \\
 \dot{\theta} & \sin\theta\dot{\phi} & 0
 \end{bmatrix}\!
 \begin{bmatrix}
  s_1\\ s_2\\ s_3
 \end{bmatrix}, \label{eq:spindynamics}
\end{align}
where $\tau_{_{ex}}=\hbar/\Delta$ is defined as the exchange time. Eq.~\eqref{eq:spindynamics} describes the coherent spin dynamics in the local frame moving with $\bm{M}(\bm{r}_c,t)$. However, spin relaxation as a non-coherent process should also be taken into account. In real materials spin relaxation is very case dependent, but regardless of the underlying mechanism, it adds a term $-\frac1{\tau_{sf}}(\bm{s}-\bm{s}_{eq})$ to Eq.~\eqref{eq:spindynamics}, where $\tau_{sf}$ is the mean spin-flip time and $\bm{s}_{eq}=\{0,0,1(-1)\}$ is the local equilibrium spin configuration for the majority (minority) band $\mathcal{E}_{\uparrow}(\mathcal{E}_{\downarrow})$. Eq.~\eqref{eq:spindynamics} should be solved numerically in general, but an approximation can be made based upon the following considerations: the large gap $\Delta$ results in an extremely small $\tau_{_{ex}}$ (typically of the order of $10^{-14}\sim10^{-15}s$). Thus on the time scale marked by $\tau_{_{ex}}$, the change of magnetization is negligible, \textit{i.e.}, magnitudes of $\partial_t\bm{M}$ and $(\dot{\bm{r}}_c\!\cdot\!\nabla)\bm{M}$ are much smaller than $M_s/\tau_{ex}$. To this end, we define two small parameters $\varepsilon_1=\tau_{ex}\sin\theta\dot{\phi}$ and $\varepsilon_2=\tau_{ex}\dot{\theta}$ which satisfy $\sqrt{\varepsilon_1^2+\varepsilon_2^2}=\hbar|\dot{\bm{M}}|/(M_s\Delta)\ll1$. On the same time scale, variations of $\varepsilon_1$ and $\varepsilon_2$ are even higher order small quantities, thus it is a good approximation to treat $\varepsilon_1$ and $\varepsilon_2$ as constants, by which Eq.~\eqref{eq:spindynamics} becomes a set of first order differential equations with a constant coefficient matrix. As a result, it can be solved analytically. Given the initial condition $\bm{s}=\bm{s}_{eq}$, the solution of Eq.~\eqref{eq:spindynamics} for the majority band is obtained in its original form in Appendix B, which, when maintaining up to the lowest order in $\varepsilon_{1,2}$, becomes the following:
\begin{widetext}
\begin{subequations}
\label{eq:solutions}
\begin{align}
 s_1(t)&=\frac{\varepsilon_1-\xi\varepsilon_2}{1+\xi^2}-\frac{e^{-\xi\tilde{t}}}{1+\xi^2}[\varepsilon_1(\cos\tilde{t}+\xi\sin\tilde{t}) +\varepsilon_2(\sin\tilde{t}-\xi\cos\tilde{t})], \\
 s_2(t)&=-\frac{\xi\varepsilon_1+\varepsilon_2}{1+\xi^2}-\frac{e^{-\xi\tilde{t}}}{1+\xi^2}[\varepsilon_1(\sin\tilde{t}-\xi\cos\tilde{t}) -\varepsilon_2(\cos\tilde{t}+\xi\sin\tilde{t})], \\
 s_3(t)&=1+\frac{e^{-\xi\tilde{t}}}{1+\xi^2}(\varepsilon_1^2+\varepsilon_2^2)[\cos\tilde{t}+\xi\sin\tilde{t}],
\end{align}
\end{subequations}
\end{widetext}
where $\tilde{t}=t/\tau_{_{ex}}$ is the scaled time, and $\xi=\tau_{ex}/\tau_{sf}$ (this is usually known as the $\beta$ parameter in the literature).

As stated above, magnetization dynamics occurs on a time scale $T$ much larger than $\tau_{ex}$, thus the number $N=T/\tau_{ex}\gg1$. This allows us to take a time average of the electron spin by defining $\langle s_i\rangle=\frac{1}{T}\int_0^Ts_i(t)\mathrm{d}t$. Then all time dependent terms in Eq.~\eqref{eq:solutions} will be negligible, because according to the following expressions
\begin{subequations}
\label{eq:approximation}
\begin{align}
 &\frac{1}{T}\int_0^T\mathrm{d}t\ e^{-\xi\tilde{t}}\cos\tilde{t}=\frac{\xi+e^{-N\xi}(\sin N-\xi\cos N)}{N(1+\xi^2)} \notag\\
 &\qquad\qquad\qquad<\frac{1}{N}\left[\frac{\xi+\sqrt{1+\xi^2}}{1+\xi^2}\right]\le\frac{1}{N}\frac{3\sqrt{3}}{4}, \\
 &\frac{1}{T}\int_0^T\mathrm{d}t\ e^{-\xi\tilde{t}}\sin\tilde{t}=\frac{1-e^{-N\xi}(\xi\sin N+\cos N)}{N(1+\xi^2)} \notag\\
 &\qquad\qquad\qquad<\frac{1}{N}\left[\frac{1+\sqrt{1+\xi^2}}{1+\xi^2}\right]\le\frac{2}{N},
\end{align}
\end{subequations}
no matter how large $\xi$ is, their upper bounds are suppressed by $N\gg1$. Thus only time-independent terms of Eq.~\eqref{eq:solutions} survive after the time averaging:
\begin{subequations}
\label{eq:averagespin}
\begin{align}
 \langle s_1\rangle&=\frac{\varepsilon_1-\xi\varepsilon_2}{1+\xi^2}, \\
 \langle s_2\rangle&=-\frac{\xi\varepsilon_1+\varepsilon_2}{1+\xi^2}, \\
 \langle s_3\rangle&=1.
\end{align}
\end{subequations}
If we write the spin as $\bm{s}=\bm{s}_{eq}+\delta\bm{s}$, then $\delta\bm{s}=\langle s_1\rangle\hat{e}_\theta+\langle s_2\rangle\hat{e}_\phi$. For the minority band, Eq.~\eqref{eq:averagespin} only differs by an overall minus sign. To express $\delta\bm{s}$ in terms of gauge invariant quantities, we need to make a coordinate transformation which corresponds to a rotation of basis in the tangential plane depicted in Fig.~\ref{Fig:Spin}
\begin{align}
 \begin{bmatrix}
 \dot{\bm{n}} \\ \bm{n}\times\dot{\bm{n}}
 \end{bmatrix}
 =\frac{\Omega}{\tau_{ex}}
 \begin{bmatrix}
 \varepsilon_2 & \varepsilon_1 \\
 -\varepsilon_1 & \varepsilon_2
 \end{bmatrix}
 \begin{bmatrix}
 \hat{e}_\theta \\ \hat{e}_\phi
 \end{bmatrix},
\end{align}
where $\Omega=|\dot{\bm{n}}|$. Then we obtain
\begin{align}
 \delta\bm{s}_{\uparrow,\downarrow}&=\mp\frac{\tau_{ex}}{1+\xi^2}[\bm{n}\times\dot{\bm{n}}+\xi\dot{\bm{n}}] \notag\\
 &=\mp\frac{\tau_{ex}}{1+\xi^2}[\bm{n}\times\frac{\partial\bm{n}}{\partial t}+\xi\frac{\partial\bm{n}}{\partial t} \notag\\
 &\qquad\qquad\qquad +\bm{n}\times(\dot{\bm{r}}_c\!\cdot\!\nabla)\bm{n}+\xi(\dot{\bm{r}}_c\!\cdot\!\nabla)\bm{n}], \label{eq:delspin}
\end{align}
where $\dot{\bm{n}}=\partial_t\bm{n}+(\dot{\bm{r}}_c\!\cdot\!\nabla)\bm{n}$ has been used and $\dot{\bm{r}}_c=-\frac{\partial\mathcal{E}_{\uparrow,\downarrow}}{\hbar\partial\bm{k}_c}$ is the center of mass velocity. The local non-equilibrium spin accumulation is obtained by integration
\begin{align}
 \delta\bm{m}=\mu_B\int\mathrm{d}\mathcal{E}[\mathscr{D}_{\uparrow}(\mathcal{E})g_{\uparrow}(\mathcal{E})\delta\bm{s}_{\uparrow}+\mathscr{D}_{\downarrow}(\mathcal{E})g_{\downarrow}(\mathcal{E})\delta\bm{s}_{\downarrow}], \label{eq:delm}
\end{align}
where $\mu_B$ is the Bohr magneton, $\mathscr{D}_{\uparrow,\downarrow}(\mathcal{E})$ is the density of states, and $g_{\uparrow,\downarrow}(\mathcal{E})$ represents the distribution function. In a weak electric field $\bm{E}$ and zero temperature, we have $g_{\uparrow,\downarrow}(\mathcal{E})=f_{0\uparrow,\downarrow}(\mathcal{E})+e\tau_{0\uparrow,\downarrow}\bm{E}\!\cdot\!\frac{\partial \mathcal{E}_{\uparrow,\downarrow}}{\hbar\partial\bm{k}_c}\frac{\partial f_{0\uparrow,\downarrow}}{\partial \mathcal{E}}$ where $f_{0\uparrow,\downarrow}(\mathcal{E})$ is the Fermi distribution function without electric field and $\tau_{0\uparrow,\downarrow}$ is the relaxation time. It should be noted that when the mean spin-flip time $\tau_{sf}$ is assumed to be independent of energy, it is equivalent to introducing it either in solving the Boltzmann equation or in Eq.~\eqref{eq:spindynamics}, and we have chosen the latter. Our target now is to relate $\delta\bm{m}$ to the charge current
\begin{align}
 \bm{j}_e=-\frac{e}{\hbar}\int\delta\mathcal{E}\left[\mathscr{D}_{\uparrow}(\mathcal{E})g_{\uparrow}(\mathcal{E})\frac{\partial\mathcal{E}_{\uparrow}}{\partial\bm{k}_c} +\mathscr{D}_{\downarrow}(\mathcal{E})g_{\downarrow}(\mathcal{E})\frac{\partial\mathcal{E}_{\downarrow}}{\partial\bm{k}_c}\right]. \notag
\end{align}
Regarding Eq.~\eqref{eq:delspin} and Eq.~\eqref{eq:delm}, terms involving electric field $\bm{E}$ and $\tau_{0\uparrow,\downarrow}$ can be expressed in terms of $\bm{j}_e$. After some simple algebra, we obtain
\begin{align}
 \delta\bm{m}&=\frac{\tau_{ex}}{1+\xi^2}\left[-\frac{n_0}{M_s^2}\bm{M}\times\frac{\partial\bm{M}}{\partial t}-\frac{\xi n_0}{M_s}\frac{\partial\bm{M}}{\partial t}\right. \notag\\
 &\left.+\frac{\mu_B P}{eM_s^2}\bm{M}\times (\bm{j}_e\cdot\nabla)\bm{M}+\frac{\xi\mu_B P}{eM_s}(\bm{j}_e\cdot\nabla)\bm{M}\right], \label{eq:centralresult}
\end{align}
where $P=(n_{\uparrow}^{F}-{n_\downarrow}^{F})/(n_{\uparrow}^{F}+{n_\downarrow}^{F})$ is the spin polarization with $n_{\uparrow(\downarrow)}^{F}$ being the electron density of the two bands at the Fermi level, and $n_0=\mu_B\int\mathrm{d}\mathcal{E}[\mathscr{D}_{\uparrow}(\mathcal{E})f_{0\uparrow}(\mathcal{E})-\mathscr{D}_{\downarrow}(\mathcal{E})f_{0\downarrow}(\mathcal{E})]$ is the local equilibrium spin density of conduction electrons, which represents the \emph{s}-band contribution to the total magnetization. For the \emph{s-d} model, the magnetization is mainly attributed to the \emph{d}-band electrons, thus the ratio $n_0/M_s$ should be very small. For example, in typical ferromagnetic metals (Fe, Co, Ni and their alloys), $n_0/M_s\sim10^{-2}$. Eq.~\eqref{eq:centralresult} reproduces Eq.~(8) in Ref.~[\onlinecite{ref:STT}], but the above derivation is purely microscopic, and the four terms of Eq.~\eqref{eq:centralresult} can be traced back to the four terms in Eq.~\eqref{eq:delspin}, respectively.

From Eq.~\eqref{eq:Hamiltonian}, the STT exerted on the background magnetization $\bm{M}(\bm{r},t)$ is $\bm{T}=(1/\tau_{_{ex}}M_s)\delta\bm{m}\times\bm{M}$, which should be added to the Landau-Lifshitz-Gilbert equation: $\partial\bm{M}/\partial t=\gamma\mathrm{\bm{H}}_{_{eff}}\times\bm{M}+(\alpha/M_s)\bm{M}\times\partial\bm{M}/\partial t+\bm{T}$, where $\gamma$ is the gyromagnetic ratio, $\mathrm{\bm{H}}_{_{eff}}$ is the effective magnetic field, and $\alpha$ is the Gilbert damping parameter. The final form of magnetization dynamics becomes
\begin{align}
 \frac{\partial\bm{M}}{\partial t}&=\tilde{\gamma}\mathrm{\bm{H}}_{_{eff}}\times\bm{M}+\frac{\tilde{\alpha}}{M_s}\bm{M}\times\frac{\partial\bm{M}}{\partial t} \notag\\
 &\quad+\frac{1}{1+\eta}\left[(\bm{u}\cdot\nabla)\bm{M}-\xi\frac{\bm{M}}{M_s}\times(\bm{u}\cdot\nabla)\bm{M}\right], \label{eq:finalresult}
\end{align}
where $\bm{u}=P\bm{j}_e\mu_B/eM_s(1+\xi^2)$ is the effective electron velocity, and $\eta=(n_0/M_s)/(1+\xi^2)$ is a dimensionless factor. The renormalized gyromagnetic ratio and Gilbert damping parameter are
\begin{align}
 \tilde{\gamma}=\frac{\gamma}{1+\eta},\quad\tilde{\alpha}=\frac1{1+\eta}[\alpha+\eta\xi], \label{eq:renormalize}
\end{align}
where the renormalization originates from the first two terms of Eq.~\eqref{eq:delspin} (or Eq.~\eqref{eq:centralresult}), and they are determined by the local equilibrium spin density $n_0$ which exists even in the absence of current. Eqs.~\eqref{eq:finalresult} and~\eqref{eq:renormalize} confirm the results of previous macroscopic theory~\cite{ref:STT}.

Our microscopic derivation relies on two assumptions: local equilibrium can be defined, and $\bm{M}$ is nearly constant on the time scale marked by $\tau_{ex}$. The former requires diffusive transport which is usually the case in transition metals and their alloys; the latter, however, is only true when the characteristic length of the texture $l$ (\textit{e.g.}, the domain wall width) satisfies $l\gg v_F\tau_{ex}$ where $v_F$ is the Fermi velocity, otherwise the solution Eqs.~\eqref{eq:solutions} and~\eqref{eq:averagespin} are invalid. In a recent experiment~\cite{ref:NarrowDW}, people measured the non-adiabatic torque on very narrow domain walls ($1\sim10\mathrm{nm}$) and found disagreement with Eq.~\eqref{eq:finalresult}. A rough estimate using $v_F\sim3\times10^5\mathrm{m/s}$ and $\Delta\sim1\mathrm{eV}$ tells us that $v_F\tau_{ex}$ is of the order of many angstroms, thus a domain wall of a few $\mathrm{nm}$ wide cannot be considered as $l\gg v_F\tau_{ex}$. In that case, our local solution is no longer a good approximation, because the time-dependent terms in Eq.~\eqref{eq:solutions} become important and the averaging in Eq.~\eqref{eq:averagespin} is no longer good. As a result, STT may exhibit non-local behavior and also oscillatory patterns in space.

The parameter $\xi$ determines the relative strength of the non-adiabatic torque with respect to the adiabatic torque. It is very material dependent and tunable in many different ways~\cite{ref:Miron,ref:Lepadatu}. But according to Eq.~\eqref{eq:approximation} and Eq.~\eqref{eq:averagespin}, the result is valid \textit{regardless} of the value of $\xi$; only $N=T/\tau_{ex}\gg1$ is sufficient to guarantee the negligence of the time dependent terms of Eq.~\eqref{eq:solutions}. This can be used to explain a recent experiment in which $\xi$ is as large as $1$~\cite{ref:Miron}, while the observed domain wall velocity is still fitted using the form of Eq.~\eqref{eq:finalresult}. However, we should mention that large $\xi$ is usually accompanied by large spin-orbit coupling, which brings about spin-orbit torque in addition to the non-adiabatic torque~\cite{ref:SOTorque1,ref:SOTorque2}. This is an important issue that draws people's attention very recently, but goes beyond the scope of this paper.

In another experiment, $\xi$ is enhanced by increasing impurity doping (which decreases $\tau_{sf}$), but the damping is basically not affected~\cite{ref:Lepadatu}. This can be easily understood through Eq.~\eqref{eq:renormalize}: since $n_0/M_s\sim10^{-2}$ is very small within the \emph{s-d} model description, $\eta$ is a small quantity, hence $\tilde{\alpha}$ could only be slightly renormalized even if $\xi$ has a sizable change.

A final remark concerns the spin motive force~\cite{ref:Shengyuan} $\bm{E}_{SMF}=\frac{\hbar}{2e}\bm{n}\cdot(\partial_t\bm{n}\times\nabla\bm{n})$, which is small but should be taken into consideration in a strict sense. As a result, the electric field should be replaced by the effective field $\bm{E}_{eff}=\bm{E}+\bm{E}_{SMF}$ in deriving Eq.~\eqref{eq:centralresult} from Eqs.~\eqref{eq:delspin} and~\eqref{eq:delm}. This creates an additional contribution to the renormalized $\tilde{\alpha}$, which has been studied recently via a quite different route~\cite{ref:GenDamping}.

We thank Maxim Tsoi, Elaine Li, Allan MacDonald, Xiao Li, Gregory Fiete, and Karin Everschor for helpful discussions. This work is supported by DOE (DE-FG03-02ER45958, Division of Materials Science and Engineering), the MOST Project of China (2012CB921300), NSFC (91121004), and the Welch Foundation (F-1255).

\appendix
\section{}

Set $|u\rangle=c_a|\!\uparrow\!(\bm{r}_c,t)\rangle+c_b|\!\downarrow\!(\bm{r}_c,t)\rangle$, the wave packet is $|W\rangle=\!\int\!\mathrm{d}^3\bm{k}w(\bm{k})e^{i\bm{k}\cdot\bm{r}}|\bm{k}\rangle|u\rangle$, where $w(\bm{k})$ is the profile function satisfying two conditions: $\int\mathrm{d}\bm{k}|w(\bm{k})|^2=1$ and $\int\mathrm{d}\bm{k}\bm{k}|w(\bm{k})|^2=\bm{k}_c$ with $\bm{k}_c$ being the center of mass momentum. Then following a quite standard procedure~\cite{ref:Niu}, the effective Lagrangian becomes
\begin{align}
 L=i\hbar\langle u|\frac{\mathrm{d}u}{\mathrm{d}t}\rangle+\hbar\bm{k}_c\cdot\dot{\bm{r}}_c-\langle u|H_{ex}|u\rangle.
\end{align}
Due to the orthogonality $\langle\uparrow|\downarrow\rangle=0$, the energy term becomes $\langle u|H_{ex}|u\rangle=|c_a|^2\mathcal{E}_{\uparrow}+|c_b|^2\mathcal{E}_{\downarrow}$. From Eq.~\eqref{eq:localspin}, we known that $s_3=|c_a|^2-|c_b|^2$ and $|c_a|^2+|c_b|^2=1$, thus we have the following:
\begin{align}
 \langle u|H_{ex}|u\rangle &=\frac{1+s_3}2\mathcal{E}_{\uparrow}+\frac{1-s_3}2\mathcal{E}_{\downarrow} \notag\\ &=\frac{\mathcal{E}_{\uparrow}+\mathcal{E}_{\downarrow}}2+s_3\frac{\mathcal{E}_{\uparrow}-\mathcal{E}_{\downarrow}}2=\mathcal{E}_0+\frac12s_3\Delta. \label{eq:energy}
\end{align}
To compute the Berry connection term, we notice that
\begin{align}
 |\frac{\mathrm{d}u}{\mathrm{d}t}\rangle=&\dot{c}_a|\uparrow\rangle+\dot{c}_b|\downarrow\rangle \notag\\
 &+\left[c_a(\dot{\bm{r}}_c\!\cdot\!\nabla+\partial_t)|\uparrow\rangle+c_b(\dot{\bm{r}}_c\!\cdot\!\nabla+\partial_t)|\downarrow\rangle\right],
\end{align}
where $\nabla=\frac{\partial}{\partial\bm{r}_c}$. Multiply by $\langle u|$ we have
\begin{align}
 \langle &u|\frac{\mathrm{d}u}{\mathrm{d}t}\rangle=(c_a^*\dot{c}_a+c_b^*\dot{c}_b) \notag\\
  &+|c_a|^2\langle\uparrow|\dot{\bm{r}}_c\!\cdot\!\nabla +\partial_t|\uparrow\rangle+c_a^*c_b\langle\uparrow|\dot{\bm{r}}_c\!\cdot\!\nabla+\partial_t|\downarrow\rangle \notag\\
 &+|c_b|^2\langle\downarrow|\dot{\bm{r}}_c\!\cdot\!\nabla+\partial_t|\downarrow\rangle +c_ac_b^*\langle\downarrow|\dot{\bm{r}}_c\!\cdot\!\nabla+\partial_t|\uparrow\rangle.
 \label{eq:udu}
\end{align}
Now define the Berry connection ($2\times2$) matrices
\begin{align}
 \bm{A}(\bm{r}_c,t)=i\hbar
 \begin{bmatrix}
 \langle\ \uparrow|\nabla|\uparrow\ \rangle && \langle\ \uparrow|\nabla|\downarrow\ \rangle \\
 \langle\ \downarrow|\nabla|\uparrow\ \rangle && \langle\ \downarrow|\nabla|\downarrow\ \rangle
 \end{bmatrix}, \label{eq:connA}
 \\
 \Phi(\bm{r}_c,t)=i\hbar
 \begin{bmatrix}
 \langle\ \uparrow|\partial_t|\uparrow\ \rangle && \langle\ \uparrow|\partial_t|\downarrow\ \rangle \\
 \langle\ \downarrow|\partial_t|\uparrow\ \rangle && \langle\ \downarrow|\partial_t|\downarrow\ \rangle
 \end{bmatrix},
 \label{eq:connPhi}
\end{align}
which play the roles of a vector potential and a scalar potential, respectively. From Eqs.~\eqref{eq:energy}, \eqref{eq:udu}, \eqref{eq:connA}, and~\eqref{eq:connPhi} we obtain the effective Lagrangian,
\begin{align}
 L=i\hbar\eta^{\dagger}\dot{\eta}+\eta^\dagger[\dot{\bm{r}}_c\!\cdot\!\bm{A}+\Phi]\eta+\hbar\bm{k}_c\!\cdot\!\dot{\bm{r}}_c-\frac12s_3\Delta-\mathcal{E}_0,
\end{align}
where $\eta=[c_a,c_b]^{\mathrm{T}}$, thus Eq.~\eqref{eq:Lagrangian} is justified. The local spin wave functions are chosen to be
\begin{align}
 |\uparrow\ \rangle=
 \begin{bmatrix}
  e^{-i\frac{\phi}2}\cos\frac{\theta}2\\
  e^{i\frac{\phi}2}\sin\frac{\theta}2
 \end{bmatrix},\qquad
 |\downarrow\ \rangle=
 \begin{bmatrix}
 -e^{-i\frac{\phi}2}\sin\frac{\theta}2\\
  e^{i\frac{\phi}2}\cos\frac{\theta}2
 \end{bmatrix},
 \label{eq:spinwavefunctions}
\end{align}
where $\theta$ and $\phi$ are spherical angles specifying the direction of local magnetization $\bm{M}(\bm{r},t)$, hence they are functions of space and time. Using Eq.~\eqref{eq:spinwavefunctions}, the Berry connections~\eqref{eq:connA} and~\eqref{eq:connPhi} can be written in a unified $2\times2$ matrix,
\begin{align}
 \mathscr{A}(\bm{r}_c,t)&\equiv\dot{\bm{r}}_c\!\cdot\!\bm{A}(\bm{r}_c,t)+\Phi(\bm{r}_c,t) \notag\\
 &=\frac{\hbar}2
 \begin{bmatrix}
 \cos\theta\dot{\phi} && -\sin\theta\dot{\phi}-i\dot{\theta}\ \\
 -\sin\theta\dot{\phi}+i\dot{\theta} && -\cos\theta\dot{\phi}
 \end{bmatrix},
\end{align}
where $\dot{\theta}=\dot{\bm{r}}_c\cdot\nabla\theta+\partial_t\theta$ and $\dot{\phi}=\dot{\bm{r}}_c\cdot\nabla\phi+\partial_t\phi$ are total time derivatives. It should be noted that the choice of Eq.~\eqref{eq:spinwavefunctions} is not unique, which gives rise to the gauge freedom of the Berry potential.

\section{}

Decomposing the Berry potential $\mathscr{A}$ in terms of Pauli matrices $\mathscr{A}=\sigma_i\mathcal{A}_i$ (adjoint representation), we have
\begin{align}
\{\mathcal{A}_1,\mathcal{A}_2,\mathcal{A}_3\}= \frac12\mathrm{Tr}[\bm{\sigma}\mathscr{A}]=\frac12\{-\sin\theta\dot{\phi},\ \dot{\theta},\ \cos\theta\dot{\phi}\}, \label{eq:adjointA}
\end{align}
where $\mathrm{Tr}[\sigma_i\sigma_j]=2\delta_{ij}$ has been used.

Taking the variation of the Lagrangian with respect to $\eta$, we obtain the evolution of the spin wave function in the local frame,
\begin{align}
 i\hbar\dot{\eta}=
 i\hbar\frac{\mathrm{d}}{\mathrm{d}t}
 \begin{bmatrix}
  c_a \\ c_b
 \end{bmatrix}
 =-\mathscr{A}
 \begin{bmatrix}
  c_a \\ c_b
 \end{bmatrix}+\frac{\Delta}2
 \begin{bmatrix}
  c_a \\ -c_b
 \end{bmatrix}. \label{eq:etadyn}
\end{align}
From Eq.~\eqref{eq:etadyn} and its complex conjugate, we derive spin dynamics in the local frame
\begin{align}
 i\hbar\frac{\mathrm{d}}{\mathrm{d}t}\bm{s}&=i\hbar\frac{\mathrm{d}}{\mathrm{d}t}(\eta^\dagger\bm{\sigma}\eta) =i\hbar(\dot{\eta}^\dagger\bm{\sigma}\eta+\eta^\dagger\bm{\sigma}\dot{\eta}) \notag\\
 &=(\eta^\dagger\mathscr{A}\bm{\sigma}\eta-\eta^\dagger\bm{\sigma}\mathscr{A}\eta) \notag\\
  &\qquad+\frac{\Delta}{2}\left([-c_a^*,c_b^*]\bm{\sigma}\eta+\eta^\dagger\bm{\sigma}
 \begin{bmatrix}
 c_a \\ -c_b
 \end{bmatrix}\right)
 \label{eq:ds}.
\end{align}
To put Eq.~\eqref{eq:ds} into a simple and elegant form, we should write it down component by component. The third component of Eq.~\eqref{eq:ds} reads:
\begin{align}
 i\hbar\dot{s}_3&=\eta^\dagger \mathcal{A}_i[\sigma_i,\sigma_3] \eta+\frac{\Delta}{2}(-|c_a|^2-|c_b|^2+|c_a|^2+|c_b|^2) \notag\\
 &=-2i\hbar\eta \varepsilon_{3ij}\mathcal{A}_i\sigma_j \eta+0 = 2i\hbar \varepsilon_{3ij}s_i\mathcal{A}_j, \label{s3}
\end{align}
where $\varepsilon_{ijk}$ is the total antisymmetric tensor. The first component reads:
\begin{align}
 i\hbar\dot{s}_1&=\eta^\dagger \mathcal{A}_i[\sigma_i,\sigma_1] \eta+\Delta(c_ac_b^*-c_a^*c_b) \notag\\
 &=-2i\hbar\eta \varepsilon_{1ij}\mathcal{A}_i\sigma_j \eta+2i\Delta\mathrm{Im}[c_ac_b^*] \notag\\
 &= 2i\hbar\varepsilon_{1ij} s_i\mathcal{A}_j-i\Delta s_2, \label{s1}
\end{align}
and the second component reads:
\begin{align}
 i\hbar\dot{s}_2&=\eta^\dagger \mathcal{A}_i[\sigma_i,\sigma_2] \eta+i\Delta(c_ac_b^*+c_a^*c_b) \notag\\
 &=-2i\hbar\eta \varepsilon_{2ij}\mathcal{A}_i\sigma_j \eta+2i\Delta\mathrm{Re}[c_ac_b^*] \notag\\
 &= 2i\hbar\varepsilon_{2ij} s_i\mathcal{A}_j +i\Delta s_1. \label{s2}
\end{align}
Now we are able to combine Eqs.~\eqref{s3},~\eqref{s1},~\eqref{s2} in a matrix form:
\begin{align}
 \begin{bmatrix}
  \dot{s}_1\\ \dot{s}_2\\ \dot{s}_3
 \end{bmatrix}
\!=\!\begin{bmatrix}
 0 & \cos\theta\dot{\phi}-\dfrac{\Delta}{\hbar} & -\dot{\theta} \\
 -\cos\theta\dot{\phi}+\dfrac{\Delta}{\hbar} & 0 & -\sin\theta\dot{\phi} \\
 \dot{\theta} & \sin\theta\dot{\phi} & 0
 \end{bmatrix}\!
 \begin{bmatrix}
  s_1\\ s_2\\ s_3
 \end{bmatrix}, \label{eq:derivecentral}
\end{align}
where Eq.~\eqref{eq:adjointA} has been used. Define $\tau_{ex}=\hbar/\Delta$ as the exchange time, Eq.~\eqref{eq:spindynamics} is justified.

As $\sin\theta\dot{\phi}$, $\cos\theta\dot{\phi}$, and $\dot{\theta}$ can be treated as constants on the time scale marked by $\tau_{ex}$, Eq.~\eqref{eq:derivecentral} can be solved analytically. Adding the relaxation term, the solution is obtained upon the initial condition $\bm{s}=\bm{s}_{eq}=\{0,0,1\}$ for the majority band,
\begin{widetext}
\begin{subequations}
\begin{align}
 s_1(t)&=\frac1{\Omega^2+1/\tau_{sf}^2}\left\{ \frac1{\tau}\sin\theta\dot{\phi}-\frac1{\tau_{sf}}\dot{\theta}-e^{-t/\tau_{sf}}\left[ \frac1\tau\sin\theta\dot{\phi}\left(\cos\Omega t+\frac1{\Omega\tau_{sf}}\sin\Omega t\right)+\Omega\dot{\theta}\left(\sin\Omega t-\frac1{\Omega\tau_{sf}}\cos\Omega t \right) \right] \right\}, \notag\\
 s_2(t)&=\frac{-1}{\Omega^2+1/\tau_{sf}^2}\left\{ \frac1{\tau}\dot{\theta}+\frac1{\tau_{sf}}\sin\theta\dot{\phi}+e^{-t/\tau_{sf}}\left[ \Omega\sin\theta\dot{\phi}\left(\sin\Omega t-\frac1{\Omega\tau_{sf}}\cos\Omega t\right)-\frac1\tau\dot{\theta}\left(\cos\Omega t+\frac1{\Omega\tau_{sf}}\sin\Omega t \right) \right] \right\}, \notag\\
 s_3(t)&=\frac1{\Omega^2+1/\tau_{sf}^2}\left\{ \frac1{\tau^2}+\frac1{\tau_{sf}^2}+e^{-t/\tau_{sf}}(\sin\theta\dot{\phi}^2+\dot{\theta}^2)\left(  \cos\Omega t+\frac1{\Omega\tau_{sf}}\sin\Omega t \right) \right\}, \notag
\end{align}
\end{subequations}
\end{widetext}
where we have defined $\Omega^2=1/\tau^2+(\sin\theta\dot{\phi}^2+\dot{\theta}^2)$ and $1/\tau=1/\tau_{ex}-\cos\theta\dot{\phi}$. Since $\varepsilon_1=\tau_{ex}\sin\theta\dot{\phi}$ and $\varepsilon_2=\tau_{ex}\dot{\theta}$ are small quantities, we have $\Omega\sim\frac1{\tau_{ex}}[1+\mathcal{O}(\varepsilon^2)]$ where the first order terms $\mathcal{O}(\varepsilon)$ all vanish. Regarding this, we neglect second order terms $\mathcal{O}(\varepsilon^2)$ in the above equations, by which Eq.~\eqref{eq:solutions} is justified.

\end{document}